\newcommand{\rem}[1]{}
\documentclass{amsart}
\usepackage{amsfonts,amssymb,amsmath,amsthm,mathrsfs}
\usepackage{url}
\usepackage[dvips]{epsfig}
\newtheorem{thrm}{Theorem}[section]
\newtheorem{lem}[thrm]{Lemma}

\newtheorem{cor}[thrm]{Corollary}

\newtheorem{remark}[thrm]{Remark}
\theoremstyle{definition}

\begin{document}
\author[C.~A.~Mantica and L.~G.~Molinari]
{Carlo~Alberto~Mantica and Luca~Guido~Molinari}
\address{C.~A.~Mantica: I.I.S. Lagrange, Via L. Modignani 65, 
20161, Milano, Italy -- L.~G.~Molinari (corresponding author): Physics Department,
Universit\`a degli Studi di Milano and I.N.F.N. sez. Milano,
Via Celoria 16, 20133 Milano, Italy.}
\email{carloalberto.mantica@libero.it, luca.molinari@unimi.it}
\subjclass[2010]{Primary 53B30, 53B50, Secondary 53C80, 83C15}
\keywords{shear, vorticity, perfect fluid, shear-free conjecture, torse-forming vector, Generalized Robertson-Walker spacetimes}
\title[Shear and vorticity of perfect fluid spacetimes]
{Shear and vorticity of perfect-fluid spacetimes\\
and the shear-free conjecture }
\begin{abstract} 
We obtain expressions for the shear and the vorticity tensors of perfect-fluid spacetimes, in terms of  
the divergence of the Weyl tensor. For such spacetimes, we prove that if the gradient of the energy density is parallel to the velocity, then
either the expansion rate is zero, or the vorticity vanishes. This statement recalls the ``shear-free conjecture'' for a perfect barotropic fluid: vanishing shear implies either vanishing expansion rate or vanishing vorticity. 
Finally, we give a new condition for a perfect fluid to be a Generalized Robinson-Walker spacetime.
\end{abstract}
\date{11 sept 2017}
\maketitle
\section{Introduction}
 
In a Lorentzian manifold, a velocity field $u^k$ is a time-like unit vector field, $u_ku^k=-1$. Its covariant gradient
has a standard decomposition into terms with geometric meaning \cite{HawkingEllis}:
\begin{align}
\nabla_k u_l = \tfrac{\theta}{n-1}h_{kl} - u_k\dot u_l+ \sigma_{kl}+\omega_{kl}   \label{0.1}
\end{align}
Namely, $\theta = \nabla_k u^k$ is the expansion scalar, $h_{kl}=g_{kl}+u_ku_l$ is an orthogonal projector,  $\dot u_k = u^m\nabla_m u_k$ is the acceleration (orthogonal to the velocity),  
\begin{align}
\omega_{kl} =& \tfrac{1}{2}h_k{}^i h_l{}^j(\nabla_iu_j - \nabla_j u_i)  \nonumber \\
=& \tfrac{1}{2}(\nabla_ku_l - \nabla_l u_k) + \tfrac{1}{2}(u_k\dot u_l - u_l \dot u_k)  \label{1.16}
\end{align}
is the (antisymmetric) vorticity tensor, with $\omega_{jk}u^k=0$,  and
\begin{align}
\sigma_{kl} =& \tfrac{1}{2}h_k{}^i h_l{}^j(\nabla_iu_j + \nabla_j u_i) - \tfrac{\theta}{n-1} h_{kl} \nonumber \\
=& \tfrac{1}{2}(\nabla_ku_l +\nabla_l u_k-\tfrac{2\theta}{n-1}h_{kl}) + \tfrac{1}{2}(u_k\dot u_l + u_l \dot u_k)  
\end{align}
is the (traceless, symmetric) shear tensor, with  $\sigma_{jk}u^k=0$. Their expressions depend on the 
connection of the spacetime. 
In this paper we consider perfect-fluid spacetimes, i.e. Lorentzian manifolds 
whose Ricci tensor takes the form
\begin{align}
R_{ij} = A g_{ij} + B u_i u_j    \label{0.6}
\end{align}
where $u_k$ is a velocity field and $A, B$ are scalar fields, with $B\neq 0$.
Geometers identify the special form \eqref{0.6} of the Ricci
tensor as the defining property of ``quasi Einstein manifolds" (with arbitrary metric signature).
The case $A=0$ defines  ``Ricci-simple" manifolds  \cite{ManSuhDe}.
For perfect-fluid spacetimes of dimension $n$ we obtain the following expressions of the terms in \eqref{0.1} (Theorem \ref{thrm2.1}): 
\begin{align}
& \theta = -\frac{1}{2B}u^k\nabla_k [ (n-2)A + B ] \label{eq1}  \\
& \dot u_j = \frac{1}{2B} h_j{}^k \nabla_k [(n-2)A-B]   \label{eq2}\\
& \sigma_{kl} = - \frac{1}{2B} \frac{n-2}{n-3} u^j ( h_k{}^q \nabla^m C_{jlqm} + h_l{}^q \nabla^m C_{jkqm}) \label{eq3}\\
& \omega_{kl} = \frac{1}{2B} \frac{n-2}{n-3} u^j ( h_k{}^q \nabla^m C_{jlqm} - h_l{}^q \nabla^m C_{jkqm}) \label{eq4}
\end{align}
where $C_{jklm}$ is Weyl's curvature tensor.
This new result enables us to prove an interesting statement (Theorem \ref{thrm_vorticity}): 

{\em If the velocity and the Weyl tensor of  a perfect-fluid
spacetime satisfy 
\begin{align}
u^j u^l \nabla^m C_{jklm}=0 \label{uunablaWeyl}
\end{align}
then either the expansion rate $\theta$ or the vorticity $\omega_{kl}$ vanish. 
The condition is equivalent to $h_k^j\nabla_j [(n-2)A+B]=0$.
In general relativity it corresponds to the gradient of the energy density being
parallel to the velocity}.

The same twofold conclusion appears in the ``shear-free conjecture'' standing for many years:
 {\em If the velocity of a barotropic perfect fluid is shear-free $(\sigma_{ij}=0)$
then either the expansion $\theta$ or the vorticity $\omega_{ij}$ of the fluid vanishes}.
Here, the shear-free and barotropic hypotheses are replaced by \eqref{uunablaWeyl}.

For the rich history of the shear-free conjecture, with its several variants of hypotheses, we refer to the review 
by Van den Bergh and Slobodeanu \cite{VandenBergh2016}, the recent work \cite{Sikhonde2017}, 
or the book \cite{Stephani} on exact solutions of Einstein's equations (Ch. 6.2).\\
We mention some achievements. 
In 1950 G\"odel claimed that a spatially homogeneous shear-free dust model could either expand
or rotate, but not both \cite{Godel}. The condition of homogeneity was relaxed in
the ``dust shear-free theorem'' proved by Ellis with tetrads: if a dust solution of the Einstein field equations 
is shear-free in a domain, it cannot both expand and rotate in the domain \cite{Ellis1967}.
Dust matter has zero pressure. In barotropic perfect fluids, pressure and energy density are related by an equation of state 
$p = p( \rho)$ with $p+\rho \neq 0$. Collins classified the barotropic shear-free perfect fluids with $\theta \neq 0$
and null vorticity \cite{CollinsCan}.\\
Some cases where the conjecture was proven: Lang and Collins \cite{LangCollins} with the assumption that the expansion $\theta $ and the energy 
density are functionally dependent, Senovilla et al. \cite{Senovilla97} with the condition that
the acceleration is proportional to the vorticity vector (including the case $\dot u_k=0$), Sopuerta \cite{Sopuerta98} when $\theta $ and the rotation
scalar are functionally dependent.\\
In Einstein spaces, if the Weyl tensor is purely electric or magnetic, a shear-free velocity field is necessarily irrotational
unless the space-time has constant curvature \cite{Barnes1984}. The result was extended to dissipative axially-symmetric fluids \cite{Herrera},
with the finding that it is irrotational if and only if the Weyl tensor is purely electric. \\
For rotating shear-free barotropic perfect fluids, if the Weyl tensor is purely electric, then $\theta =0$ \cite{Collins1983}.
Collins and Wainwright \cite{CollinsWain} proved that any barotropic perfect fluid solution of the field equations of general relativity, which is shear-free, irrotational with $\theta\neq 0$, is either a Friedmann-Robertson-Walker model, or spherically symmetric Wyman solutions, or a special class of plane-symmetric models. In particular, if $p=(\gamma -1)\rho $, then only the FRW model remains.\\

In Section 2, for perfect fluid spacetimes, we express the components of the gradients of the velocity in terms of the 
Weyl tensor. In Section 3 we prove our main statement, that replaces the shear-free and barotropic hypotheses with the condition
\eqref{uunablaWeyl} on the Weyl tensor.
In Section 4 we discuss the structure of spacetimes admitting a velocity field that is both shear-free and vorticity-free.
In particular, in Theorem \ref{3.2} we significantly weaken the hypotheses given in ref.\cite{ManMolDe}, for a perfect fluid to
be a Generalized Robinson-Walker space-time.

\section{Velocity terms for perfect-fluid spacetimes}
\begin{thrm}\label{thrm2.1}
For a perfect-fluid spacetime with velocity field $u_j$, the expansion rate, the acceleration, the shear and vorticity tensors are given by the expressions \eqref{eq1}-\eqref{eq4}. 
\begin{proof}
The covariant divergence of eq.\eqref{0.6} is
$\nabla_j [(n-2)A-B]  = 2u_j \dot B  + 2B (\dot u_j + \theta u_j ) $, where
$\dot B=u^k\nabla_k B$;
contraction with $u^j$ gives the equation for $\theta$:  
$(n-2)\dot A +\dot B = - 2\theta B$. Elimination or $\theta$ gives the acceleration $\dot u_j$:
\begin{align}
\nabla_j [(n-2)A-B] + u_j [(n-2)\dot A -\dot B] = 2B \dot u_j 
 \end{align}
In the general expression for the divergence of the conformal tensor,
\begin{align*}
\nabla^m C_{jklm} = \frac{n-3}{n-2} \left [ \nabla_k R_{jl}-\nabla_j R_{kl} -\frac{1}{2(n-1)}(g_{jl}\nabla_k R - g_{kl}\nabla_j R)\right ]
\end{align*}
the expressions \eqref{0.6} of $R_{ij}$ and the curvature scalar $R=nA-B$ are inserted:
\begin{align*}
\tfrac{n-2}{n-3}\nabla^m C_{jklm} =  \nabla_k (Bu_j u_l) -\nabla_j (Bu_k u_l)  + (g_{jl}\nabla_k - g_{kl}\nabla_j ) \tfrac{(n-2)A+B}{2(n-1)} \nonumber
\end{align*}
The expression is contracted with $u^j$:
\begin{align*}
 \tfrac{n-2}{n-3}u^j \nabla^m C_{jklm} =&   -B ( \nabla_k u_l + u_k \dot u_l+ u_l\dot u_k)  -u_l ( \nabla_k B + u_k \dot B)  \\
&+ (u_l\nabla_k - g_{kl}u^j\nabla_j ) \tfrac{(n-2)A+B}{2(n-1)} \nonumber  
\end{align*}
The expression \eqref{0.1} for $\nabla_k u_l$ is used:
\begin{align*}
 \tfrac{n-2}{n-3}u^j \nabla^m C_{jklm} = \tfrac{(n-2)\dot A +\dot B}{2(n-1)} h_{kl}  -B(\sigma_{kl}+\omega_{kl} - u_l\dot u_k)   
  -u_l ( \nabla_k B + u_k \dot B) \\ + (u_l\nabla_k - g_{kl}u^j\nabla_j ) \tfrac{(n-2)A+B}{2(n-1)} \nonumber  \\
  = \tfrac{(n-2)\dot A - (2n-3)\dot B}{2(n-1)} u_k u_l  -B(\sigma_{kl}+\omega_{kl} - u_l\dot u_k)   
  + u_l\nabla_k  \tfrac{(n-2)A-(2n-3)B}{2(n-1)} \nonumber  
\end{align*}
The expression for the acceleration is used:
\begin{align}
B(\sigma_{kl}+\omega_{kl}) = - \tfrac{n-2}{n-3} u^j \nabla^m C_{jklm} - \tfrac{n-2}{2(n-1)} u_l h_{km} \nabla^m 
[(n-2)A+B ] \label{before}
\end{align}
Contraction with $u^l$ gives an interesting relation:
\begin{align}
\tfrac{n-2}{2(n-1)} h_{km} \nabla^m  [(n-2)A+B] = - \tfrac{n-2}{n-3} u^j u^l\nabla^m C_{jklm} \label{simplification}
\end{align}
Therefore:
\begin{align}
B(\sigma_{kl}+\omega_{kl} ) = - \tfrac{n-2}{n-3} u^j h_l{}^q \nabla^m C_{jkqm} 
\end{align}
The symmetric and antisymmetric parts are separated.
\end{proof}
\end{thrm}
Explicit expressions for the tensors are obtained from \eqref{before} (the Bianchi identity fot the Weyl tensor is used to simplify the vorticity): 
\begin{align}
B\, \sigma_{kl} =& -\tfrac{n-2}{2(n-3)} u^j\nabla^m (C_{jklm}+C_{jlkm}) - \tfrac{n-2}{2(n-1)} u_l u_k [(n-2)\dot A+\dot B ]\\
&- \tfrac{n-2}{4(n-1)} (u_l\nabla_k + u_k\nabla_l) [(n-2)A+B ] \nonumber\\
B\, \omega_{kl} =& \tfrac{n-2}{2(n-3)} u^j\nabla^m C_{kljm}- \tfrac{n-2}{4(n-1)} (u_l\nabla_k - u_k\nabla_l) [(n-2)A+B ]
\end{align}

If Einstein's equations hold, $R_{ij}-\frac{1}{2}Rg_{ij}=8\pi T_{ij}$, the Ricci tensor of a perfect fluid spacetime 
corresponds to the stress-energy tensor of a perfect fluid with same velocity field $u^k$, pressure $p$ and 
energy density $\rho$: $T_{ij}=(p+\rho) u_iu_j + p g_{ij}$. In this correspondence it is $(n-2)A-B= -16 \pi p$ 
and $(n-2)A+B=16 \pi \rho$. Therefore:
\begin{align}
\theta =& -\frac{u^k\nabla_k \rho}{\rho + p} \\
\dot u_j =& -\frac{1}{\rho+p} h_j{}^k \nabla_k p   \\
 (\rho+p)\, \sigma_{kl} =& -\tfrac{n-2}{16\pi (n-3)} u^j\nabla^m (C_{jklm}+C_{jlkm}) - \tfrac{n-2}{n-1} u_l u_k \dot\rho\\
&-  \tfrac{n-2}{2(n-1)} (u_l\nabla_k + u_k\nabla_l) \rho \nonumber\\
 (\rho+p)\, \omega_{kl} =& \tfrac{n-2}{16\pi(n-3)} u^j\nabla^m C_{kljm}- \tfrac{n-2}{2(n-1)} (u_l\nabla_k - u_k\nabla_l) \rho
\end{align}
The requirement $B\neq 0$ for perfect-fluid spacetimes means $p+\rho\neq 0$.
\section{A condition for vanishing vorticity}
If the velocity field is closed $(\nabla_i u_j =\nabla_j u_i)$, then it is geodesic $(\dot u_k =0)$ and the vorticity is zero. We now obtain a
different condition for zero vorticity. It is based on the following two lemma by Senovilla et al. \cite{Senovilla97}:
\begin{lem}\label{lemmaone}
Let  $A_{jkl} = \tfrac{1}{2}[ u_j (\nabla_k u_l-\nabla_l u_k)+u_k(\nabla_l u_j-\nabla_j u_l)+u_l(\nabla_j u_k-\nabla_k u_j)]$. 1) $\omega_{kl}=0$ if and only if $A_{jkl}=0$.\\
2) If there are scalar functions $\lambda, \,f$ such that $u_j = \lambda \nabla_j f$ then $\omega_{kl}=0$.
\begin{proof}
By the identity $u_j \omega_{kl}+u_k \omega_{lj}+u_l \omega_{jk} = A_{jkl}$ it follows that $\omega_{kl}=0$
means $A_{jkl}=0$. The other way, if $A_{jkl}=0$ then one evaluates $0=u^j A_{jkl} = \tfrac{1}{2}(-\nabla_k u_l + \nabla_l u_k -u_k \dot u_l +u_l\dot u_k) = -\omega_{kl}$.\\
If $u_j=\lambda \nabla_j f$ then direct substitution gives $A_{jkl}=0$, i.e. $\omega_{kl}=0$. 
\end{proof}
\end{lem}

\begin{lem}\label{lemmabis}
If $h_k{}^m\nabla_m f=0$ then, either $f$ is constant, or $\omega_{ij}=0$.
\begin{proof}
Being $\nabla_k f+u_k \dot f=0$, if $\dot f=0$ it follows that $\nabla_k f=0$ i.e. $f$ is constant. If $\dot f\neq 0$ then 
$u_k =-(\nabla_k f)/\dot f$. Then, by Lemma \ref{lemmaone}, $\omega_{kl}=0$.
\end{proof}
\end{lem}

Now, thanks to the expressions \eqref{eq1}-\eqref{eq4}, the theorem follows: 

\begin{thrm}\label{thrm_vorticity}
In a perfect fluid, if $u^qu^p\nabla^m C_{qkpm}=0$ then either $\theta=0$  or $\omega_{kl}=0$.
\begin{proof}
From the identity \eqref{simplification} it follows that $h_k{}^m \nabla_m [(n-2)A+B]=0$. Then
Lemma \ref{lemmabis} implies that either 
$(n-2) A+ B$ is constant in spacetime, i.e. $\theta=0$, or $\omega_{kl}=0$.
\end{proof}
\end{thrm}

\begin{remark}
In general relativity the condition in the theorem is $h_k{}^m \nabla_m \rho =0 $, meaning that the gradient of the energy density is
parallel to the velocity. The possible outcome of the theorem $\nabla_k [(n-2)A+B]=0$ means that the energy density $\rho $ is uniform in spacetime.
\end{remark}

\section{Shear and vorticity-free perfect-fluid spacetimes} 
The vanishing of terms in the decomposition \eqref{0.1} of a velocity field, determines a hierarchy of spacetimes.
If a spacetime admits a shear-free and vorticity-free velocity field, 
$\nabla_i u_j = \frac{\theta}{n-1}h_{ij}-u_i\dot u_j$, then it admits 
a line element of the form \cite{Ponge}:
\begin{align}
 ds^2 = - \varphi (t,\vec x)^2 dt^2  + f(t,\vec x)^2 g^*_{\mu\nu}(\vec x) dx^\mu dx^\nu  \label{0.2}
\end{align}
In the geometric literature, \eqref{0.2} represents a doubly twisted product.\\
If the velocity field is also geodesic, $\dot u_k=0$, then it satisfies the torse-forming condition 
$\nabla_i u_j = \tfrac{\theta}{n-1} h_{ij}$, which is necessary and sufficient for the spacetime to be twisted \cite{twisted}, \cite{ChenBook}, with metric 
\begin{align*}
ds^2 = -dt^2 + f(t,\vec x)^2 g^*_{\mu\nu} (\vec x) dx^\mu dx^\nu  
\end{align*}
If the velocity  is also an eigenvector of the Ricci tensor, the spacetime is a Generalized Robertson-Walker  (GRW) spacetime  \cite{survey}, with metric 
\begin{align*}
ds^2 = -dt^2 + f(t)^2 g^*_{\mu\nu}(\vec x) dx^\mu dx^\nu
\end{align*}
(if the Weyl tensor is zero, it is a Robertson-Walker space-time).
Finally, if also $\theta=0$ the space is the product of disjoint manifolds: $ds^2 = -dt^2+ g^*_{\mu\nu}(\vec x) dx^\mu dx^\nu$. 

As a consequence of Theorem \ref{thrm2.1} we have: 
\begin{cor}
The velocity of a perfect-fluid space-time is shear-free and vorticity-free if and only if the Weyl tensor satisfies 
$h_i{}^l u^j \nabla^m C_{jklm} = 0$ i.e.
\begin{align}
u^j( u_i u^l  \nabla^m C_{jklm} +  \nabla^m C_{jkim} ) = 0. \label{generalcond}
\end{align}
\end{cor}
The simplest case $\nabla^m C_{jklm}=0$ includes locally symmetric, conformally flat and conformally symmetric 
perfect-fluid spacetimes. 
In \cite{ManMolDe} we proved that a perfect-fluid spacetime with $\nabla_i u_j = \nabla_j u_i$ and
$\nabla_m C_{jkl}{}^m=0$ is a GRW spacetime. 
However, both hypotheses of closedness and conformal harmonicity can now be weakened: 
\begin{thrm}\label{3.2}
Let M be a perfect-fluid spacetime. If  $u^k$ is geodesic $(\dot u^k=0)$ and if
$h_i{}^l u^j \nabla_m C_{jkl}{}^m=0$, then $M$ is a GRW spacetime.
\begin{proof}
The condition on the Weyl tensor implies that the velocity is both shear-free and vorticity-free. Being geodesic, the
velocity is torse-forming: $\nabla_i u_j = \theta h_{ij}/(n-1)$. For a perfect-fluid spacetime, the velocity
is an eigenvector of the Ricci tensor. Then M is a GRW spacetime. 
\end{proof}
\end{thrm}

Now we give an example of perfect-fluid space-time that is shear-free and vorticity-free \cite{ManticaSuh}.
The Ricci tensor has the form \eqref{0.6}, and is also recurrent with 1-form $\eta_k\neq 0$:
\begin{align}
\nabla_k R_{jl}-\nabla_j R_{kl} = \eta_k R_{jl} - \eta_j R_{kl} \label{recurrence}
\end{align}
A contraction with the metric tensor gives: $\nabla_k R =2\eta_k [(n-1)A-B]-2Bu_ku^j\eta_j$. Contraction with $u^k$ gives: $u^k\nabla_k R= 2(n-1)Au^j\eta_j$. Next evaluate 
\begin{align*}
 u^j  \nabla_m C_{jkl}{}^m =&  \frac{n-3}{n-2} \left [\eta_k R_{jl} u^j -u^j\eta_j R_{kl} -\frac{1}{2(n-1)}(u_l\nabla_k R - g_{kl}u^j \nabla_j R) \right ] \\
=& -\frac{n-3}{n-1} B u_l h_{km}\eta^m 
\end{align*}
It is clear that $h_i{}^l u^j \nabla^m C_{jklm} = 0$. Then $\sigma_{kl}=0$ and $\omega_{kl}=0$.


%

\begin{thebibliography}{99}
%
\bibitem{Barnes1984}
A.~Barnes, {\em Shear-free flows of a perfect fluid}, in Classical General Relativity. Proceedings of the conference on classical (non-quantum) general relativity, London, Dec. 1983,
pp15--22;  Editors M.~A.~H.~MacCallum, J.~N.~Islam; W.~B.~ Bonnor, Cambridge University Press, 1984.
%
\bibitem{ChenBook}
B.~Y.~Chen, Differential Geometry of Warped Product Manifolds and Submanifolds (World Scientific, Singapore 2017)
%
\bibitem{Collins1983}
C.~B.~Collins, {\em Shear-free perfect fluids with zero magnetic Weyl tensor}, J. Math. Phys. {\bf 25} (4) (1984) 995--1000.
%
\bibitem{CollinsCan}
C.~B.~Collins, {\em Shear-free fluids in general relativity}, Can. J. Phys. {\bf 64} (1986) 191
%
\bibitem{CollinsWain}
C.~B.~Collins and J.~Wainwright, {\em Role of shear in general-relativistic cosmological and stellar models}, Phys. Rev. D {\bf 27} n.6 (1983) 1209--18.
%
%
\bibitem{Ellis1967}
G.~F.~R.~Ellis, {\em  Dynamics of pressure-free matter in General Relativity}, J. Math. Phys. {\bf 8} (5) (1967) 1171--1194.
%
\bibitem{HawkingEllis}
S.~W.~Hawking and G.~F.~R.~Ellis, The large scale structure of space time, Cambridge Monographs on
Mathematical Physics Vol 1 (Cambridge University Press, London, 1973).
%
\bibitem{Herrera}
L.~Herrera, A.~Di Prisco, J.~Ospino, {\em Shear-free axially symmetric dissipative fluids}, Phys.  Rev. D {\bf 89} (2014) 127502--5.
%
%
\bibitem{Godel}
K.~G\"odel, {\em Rotating universes in General Relativity Theory}, Proceedings of
the International Congress of Mathematicians, Cambridge, Mass., 1950, vol. 1, 175-
181 (A.M.S., R.I., 1952).

\bibitem{LangCollins}
J.~M.~Lang and C.~B.~Collins, {\em Observationally homogeneous shear-free perfect fluids}, Gen. Relativ. Gravit.
{\bf 20} n.7 (1988) 683--710.
%
\bibitem{ManMolDe}
C.~A.~Mantica, L.~G.~Molinari and U.~C.~De, {\em A condition for a perfect-fluid spacetime to be a generalized Robertson-Walker spacetime}, 
J. Math. Phys. 57 (2016), 022508 (6pp.); Erratum, J. Math.Phys. 57 (2016) 049901.
%
\bibitem{survey}
C.~A.~Mantica and L.~G.~Molinari, {\em Generalized Robertson-Walker spacetimes, a survey}, Int. J. Geom. Meth. Mod. Phys. {\bf 14} n.3 (2017) 1730001, 27 pp.
%
\bibitem{twisted}
C.~A.~Mantica and L.~G.~Molinari, {\em Twisted Lorentzian manifolds: a characterization with torse-forming time-like unit vectors}, Gen. Relativ. Gravit. {\bf 49} (2017) 51.
%
\bibitem{ManSuhDe}
C.~A.~Mantica, Y.~J.~Suh and U.~C.~De,
{\em A note on generalized RobertsonÐWalker space-times}, Int. J. Geom. Meth. Mod. Phys. {\bf 13} (2016)
1650079 (9pp.)
%
\bibitem{ManticaSuh}
C.~A. ~Mantica and  Y.~J.~Suh, {\em Recurrent conformal 2-forms  on pseudo-Riemannian manifolds},  Int. J. Geom. Meth. Mod. Phys.  {\bf 11} n.6  (2014) 1450056 (29pp.).
%
\bibitem{Ponge}
R.~Ponge and H.~Reckziegel, {\em Twisted products in pseudo-Riemannian geometry}, Geom. Dedicata {\bf 48} (1993) 15--25.
%
%
\bibitem{Senovilla97}
J.~M.~M.~Senovilla, C.~F.~Sopuerta and P.~Szekeres, {\em Theorems on shear-free perfect fluids with their Newtonian analogues}, Gen. Relativ. Gravit. {\bf 30} n.3 (1998) 389--411.  
%
\bibitem{Sikhonde2017}
M.~E.~Sikhonde and P.~K.~S.~Dunsby, {\em Reviving the shear-free perfect fluid conjecture in general relativity},
arXiv:1708.02462 [gr-qc].
%
\bibitem{Sopuerta98}
C.~F.~Sopuerta, {\em Covariant study of a conjecture on shear-free barotropic perfect fluids}, Class. Quantum Gravity {\bf 15} (1998) 1043--1062.
%
\bibitem{Stephani}
H.~Sthepani, D.~Kramer, M.~MacCallum, C.~Hoenselaers and E.~Hertl, Exact solutions of Einstein's Field
Equations, Cambridge Monographs on Mathematical Physics (Cambridge University Press 2nd ed., 2003).
%
\bibitem{VandenBergh2016}
N.~Van den Bergh and R.~Slobodeanu, {\em Shear-free perfect fluids with a barotropic
equation of state in general relativity: the present status}, Class. Quantum Grav. {\bf 33} (2016) 085008 (30pp)
\end{thebibliography}
\end{document}